\documentclass[12pt,nohyper,letterpaper]{JHEP3}         
\usepackage{amssymb,amsfonts,amsmath}

\usepackage{epsfig}

\def\be{\begin{eqnarray}}
\def\ee{\end{eqnarray}}

\newcommand\para{\paragraph{}}
\newcommand{\ft}[2]{{\textstyle\frac{#1}{#2}}}
\newcommand{\eqn}[1]{(\ref{#1})}

\def\Dslash{\,\,{\raise.15ex\hbox{/}\mkern-12mu D}}
\def\Dbarslash{\,\,{\raise.15ex\hbox{/}\mkern-12mu {\bar D}}}
\def\delslash{\,\,{\raise.15ex\hbox{/}\mkern-9mu \partial}}
\def\delbarslash{\,\,{\raise.15ex\hbox{/}\mkern-9mu {\bar\partial}}}
\def\pslash{\,\,{\raise.15ex\hbox{/}\mkern-9mu p}}
\def\calDslash{\,\,{\raise.15ex\hbox{/}\mkern-12mu {\cal D}}}

\newcommand{\CP}{{\bf CP}}

\def\lae{\mathrel{\mathop{\smash{\lower .5 ex \hbox{$\stackrel<\sim$}}}}}
\def\lae{\mathrel{\mathop{\smash{\lower .5 ex \hbox{$\stackrel>\sim$}}}}}

\def\spup{\left|\, \uparrow \right. \rangle}
\def\spdn{\left|\, \downarrow \right. \rangle}

\def\Dslash{\,\,{\raise.15ex\hbox{/}\mkern-13mu D}}
\def\Dbarslash{\,\,{\raise.15ex\hbox{/}\mkern-12mu {\bar D}}}
\def\delslash{\,\,{\raise.15ex\hbox{/}\mkern-10mu \partial}}
\def\delbarslash{\,\,{\raise.15ex\hbox{/}\mkern-9mu {\bar\partial}}}
\def\pslash{\,\,{\raise.15ex\hbox{/}\mkern-11mu p}}
\def\qslash{\,\,{\raise.15ex\hbox{/}\mkern-9mu q}}
     \def\kslash{\,\,{\raise.15ex\hbox{/}\mkern-11mu k}}
\def\eslash{\,\,{\raise.15ex\hbox{/}\mkern-9mu \epsilon}}
\def\calDslash{\,\,{\rais.15ex\hbox{/}\mkern-12mu {\cal D}}}

\title{Non-Abelian~Berry~Phases~and~BPS~Monopoles}
\author{Julian Sonner\footnote{On leave from DAMTP and Trinity College, Cambridge, {\tt j.sonner@damtp.cam.ac.uk}}\\
Center for Theoretical Physics, \\
Massachusetts Institute of Technology\\
Cambridge, MA 02139, USA\\
Email: {\tt sonner@mit.edu}
}
\author{David Tong\\
Department of Applied Mathematics and Theoretical Physics, \\
University of Cambridge, UK\\
Email: {\tt d.tong@damtp.cam.ac.uk}
}

\preprint{MIT-CTP 3982}
\abstract{We study a simple quantum mechanical model of a spinning particle moving on a sphere in the presence of a magnetic field. The system has two ground states. As the magnetic field is varied, the ground states mix through a non-Abelian Berry phase. We show that this Berry phase is the path ordered exponential of the smooth $SU(2)$ 't Hooft-Polyakov monopole. We further show that, by adjusting a potential on the sphere, the monopole becomes BPS and obeys the Bogomolnyi equations. For this choice of potential, it turns out that there is a hidden supersymmetry underlying the system and the Bogomolnyi equations are analogous to the tt* equations of Cecotti and Vafa. We conjecture that the Bogomolnyi equations also govern the Berry phase of ${\cal N}=(2,2)$ supersymmetric sigma models with other target spaces.}

\begin{document}
\pagestyle{plain} \setcounter{page}{1}
\newcounter{bean}
\baselineskip16pt

\section{Introduction}

In 1974, 't Hooft and Polyakov discovered a new solution of Yang-Mills-Higgs theories  \cite{thooft,polyakov}. At large distances, it looks like a Dirac magnetic monopole. However, the configuration is smooth, with the singularity at the origin of the Dirac monopole resolved by the non-Abelian gauge fields.

\para
The spatial profile of the 't Hooft-Polyakov field configuration depends on the scalar potential for the adjoint-valued Higgs field $\phi$. Among these, one profile is rather special. This occurs when the potential vanishes and, as first shown by Prasad and Sommerfield \cite{ps}, it is possible to find an exact solution. Later, Bogomolnyi \cite{bog} showed that the non-Abelian field strength, ${\cal F}_{\mu\nu}$, for this configuration solves the simple, first order, differential equations
\be {\cal F}_{\mu\nu} = \epsilon_{\mu\nu\rho}{\cal D}_\rho \phi\, \label{bog}\ee
where ${\cal F}_{\mu\nu}=\partial_\mu{\cal A}_\nu-\partial_\nu{\cal A}_{\mu}+[{\cal A}_\mu,{\cal A}_\nu]$ and ${\cal D}_\rho\phi=\partial_\rho\phi + [{\cal A}_\rho,\phi]$. Monopoles of this type are known as BPS, after the three authors named above. The subsequent discovery that these monopole play a special role in supersymmetric theories \cite{wo} has resulted in the title ``BPS" being ascribed to almost anything associated to supersymmetry.

\para
There is another, more abstract, situation in theoretical physics where the Dirac monopole arises. This is the Berry phase in quantum mechanics. Consider a spin 1/2 particle in a magnetic field $\vec{B}$. The Hamiltonian is given by,
\be H=- \vec{B}\cdot\vec{\sigma} -|\vec{B}|\,1_{2}\ , \label{ham}\ee
where $\vec{\sigma}$ are the Pauli matrices and $1_2$ is the unit $2\times 2$ matrix, whose presence in the Hamiltonian simply ensures that the ground state energy of this two-state system is normalized to zero. We start in this ground state, $|0\rangle$. We then slowly rotate the magnetic field $\vec{B}$ until, finally, we return to our initial set-up. The adiabatic theorem in quantum mechanics tells us that the system remains in the ground state and changes only by a phase. The question is: what is this phase? Since we have normalized the vacuum to zero energy, there is no dynamical contribution. Nonetheless, Berry showed that there is a geometrical phase which depends on the path $\Gamma$ taken in the space of magnetic fields \cite{berry,simon},
\be |\,0\rangle \rightarrow \,\exp\left( - i\oint_\Gamma
\vec{A}\cdot d\vec{B}\right)\,|\,0\rangle\ .\ee
The Abelian Berry connection $\vec{A}$ is defined in terms of the dependence of the ground state on the magnetic field $\vec{B}$,
\be \vec{A} = i\langle 0|\frac{\partial}{\partial
\vec{B}} |\,0\rangle \label{time}\ .\ee
Berry showed that, for the simple Hamiltonian \eqn{ham}, the connection \eqn{time} is that of the Dirac magnetic monopole: $\vec{A}=\vec{A}^{\ \rm Dirac}$. One can form a $U(1)$ field strength from the Berry connection in the usual way $F_{\mu\nu} = \frac{\partial A_\mu}{\partial B_\nu} - \frac{\partial A_\nu}{\partial B_\mu}$. This takes the radial, Dirac monopole form
\be F_{\mu\nu}  = \epsilon_{\mu\nu\rho} \frac{B_\rho}{B^3}\ .\ee
Note that there's a potential for confusion here, because $\epsilon_{\mu\nu\rho}F_{\mu\nu}$ is an abstract magnetic monopole over the space of real magnetic fields $\vec{B}$. The field strength $F_{\mu\nu}$ has a singularity at the origin. This is nothing to be afraid of: it simply reflects the fact that the excited state and the ground state become degenerate at $\vec{B}=0$. Indeed, the very existence of the Berry phase can be traced to this degenerate point in parameter space.

\para
In this paper, we ask whether the smooth non-Abelian 't Hooft-Polyakov monopole can appear as a Berry connection in simple quantum mechanical systems. The answer, as we shall see, is yes. The concept of the non-Abelian Berry connection was introduced by Wilczek and Zee \cite{wz}. This occurs if a system has degenerate eigenstates for all values of the parameters $\vec{B}$. If there are $N$ degenerate states  $|a\rangle$, $a=1,\ldots, N$, then after a cyclic and adiabatic tour through the space of parameters, the system will undergo a $U(N)$ rotation,
\be |a\rangle \rightarrow P\,\exp\left(-i\oint_\Gamma
\vec{\cal A}_{ab}\cdot d\vec{B}\right)\,|b\rangle\ ,\label{nonabb}\ee
where the $u(N)$ valued Berry connection is defined by
\be\label{eq:nonabelianberry} \vec{\cal A}_{ab} = i\langle b |\frac{\partial}{\partial
\vec{B}} |a\rangle \label{timeagain}\ .\ee
To build an $SU(2)$ 't Hooft-Polyakov monopole as a Berry connection, we need to construct a Hamiltonian  with two degenerate ground states for all values of the parameters $\vec{B}$. Moreover, since the 't Hooft-Polyakov monopole is smooth, our system should generate a topologically non-trivial Berry connection without any further degeneracies occurring in parameter space\footnote{Non-Abelian monopoles have arisen previously in the context of Berry phases \cite{berrymon,zhang}. However, in both of these papers the configuration is not smooth, with a singularity at the origin resulting from an extra degeneracy of states.}.

 \para
 In the following section, we show that these conditions arise for a spin 1/2 particle moving on a sphere ${\bf S}^2$ in the presence of a particular magnetic field $\vec{B}$. As $\vec{B}$ is varied, the mixing \eqn{nonabb} between the ground states is governed by a 't Hooft-Polyakov monopole. Moreover, we show that by including a potential over ${\bf S}^2$, the monopole takes the BPS form, and the $SU(2)$ Berry connection satisfies the Bogomolnyi equation \eqn{bog}.

\para
In Section 3 we show that, as one might suspect, when the Berry connection is the BPS monopole, there is an underlying supersymmetry. For a specific choice of potential over the sphere,  the quantum mechanical model described in Section 2 turns out to be a consistent truncation of the ${\bf CP}^1\cong {\bf S}^2$ sigma-model with ${\cal N}=(2,2)$ supersymmetry.

\para
The fact that one can write an equation, such as \eqn{bog}, to describe the Berry connection is intriguing. Typically, the only way to compute the Berry connection is through the direct definition \eqn{timeagain}, but to do this one first needs to compute the exact ground states as a function of the parameters.  If the Berry connection can be shown to obey an equation --- for example,  of the form \eqn{bog} --- then one can circumvent this step. In fact, this short-cut is known to happen in supersymmetric theories when one varies complex parameters which live in background chiral multiplets \cite{C1,C2,tt*}. In that situation, the equation obeyed by the Berry curvature is known as the tt* equation.  In contrast, the magnetic field triplet $\vec{B}$ lives in a background vector multiplet (which, in quantum mechanics, contains 3 scalars). The Bogomolnyi equations \eqn{bog} can be thought of as the analog of the tt* equations for the vector multiplet parameters in the ${\bf CP}^1$ model. At the end of Section 3, we conjecture that the same equation also describes the Berry curvature for the quantum mechanical ${\bf CP}^N$ sigma-model\footnote{{\bf Note added:} In a subsequent paper \cite{holonomy} we proved this conjecture and showed that, for a large class of supersymmetric systems, the 
Berry phase solves the Bogomolnyi equation 
\eqn{bog}, or generalizations of this equation.}.

\para
This paper is a continuation of our earlier work on understanding Berry phase in supersymmetric quantum mechanics and string theory \cite{us22,us44,d0,chris}. Applications of supersymmetric Berry phases to the microstates of black holes were considered recently in \cite{amsterdam}.

\section{Quantum Mechanics and Monopoles}\label{sec:toymodel}

In this section we introduce two simple quantum mechanical systems with degenerate ground states. Both of these have a non-Abelian $SU(2)$ Berry phase described by a 't Hooft-Polyakov monopole. For the first, we have only an implicit description of the profile of the monopole. However, a small modification of this system allows us to solve for the Berry phase exactly and we find the BPS monopole satisfying \eqn{bog}.

\subsection{A Spinning Particle on a Sphere}

Consider a neutral, spin 1/2 particle moving on a sphere ${\bf S}^2$ in the presence of a magnetic field $\vec{B}$ whose magnitude varies over the sphere. The Hamiltonian is given by
\begin{equation}
H =- \frac{\hbar^2}{2m}\,\Delta \,1_2 - \hbar\,\vec{B}\cdot\vec{\sigma} \cos\theta\ .\label{newham}
\end{equation}
The operator $\Delta$ is the Laplacian on the unit ${\bf S}^2$
\begin{equation}
\Delta = \frac{1}{\sin\theta}\frac{\partial}{\partial\theta}\left(  \sin\theta \frac{\partial}{\partial\theta}  \right) + \frac{1}{\sin^2\theta}\frac{\partial^2}{\partial\phi^2}
\ .\end{equation}
The magnetic field varies along $\theta \in (0,\pi ]$, but the azimuthal coordinate $\phi \in (0,2\pi]$ is cyclic. This ensures that the ground states of the system will not depend on $\phi$.

\para
The Hamiltonian enjoys a ${\bf Z}_2$ symmetry,
\be
\vec{B} & \rightarrow - \vec{B} \ \ \ ,\ \ \ \theta  & \rightarrow \pi - \theta\ .
\ee
The sign flip of the magnetic field acts on the Hilbert space by exchanging spin-up and spin-down states, $\spup$ and $\spdn$, defined to be the two normalized eigenvectors of $\vec{B}\cdot \vec{\sigma}$ with eigenvalues $+1$ and $-1$ respectively.

\para
The ${\bf Z}_2$ symmetry guarantees the existence of two ground states for all values of $\vec{B}$. For $\vec{B}\neq 0$, the spin-up state is localized near $\theta=0$, while the spin-down state is localized near $\theta=\pi$. When $\vec{B}=0$, both ground states are smeared uniformly over the sphere. However, in contrast to the Hamiltonian \eqn{ham}, there is no extra degeneracy of the ground states when $\vec{B}=0$. For arbitrary values of $\vec{B}$, the two, normalized, ground states are a combination of the spin states and a spatial wavefunction, $\psi(\cos\theta; B)$, which depends on the magnitude $B=|\vec{B}|$,
\be |1\rangle = \psi(\cos\theta;B)\spup\ \ \ ,\ \ \ |2\rangle=\psi(-\cos\theta;B)\spdn\ . \label{gstate}\ee
Writing $x=\cos\theta$, the spatial wavefunction $\psi(x;B)$ satisfies the Schr\"odinger  equation,
\be -\frac{\hbar^2}{2m}(1-x^2)\psi^{\prime\prime} + \frac{\hbar^2}{2m} x \psi^\prime - \hbar Bx\psi = E_0\psi\ ,\ee
with $\psi^\prime = d\psi/dx$, and $E_0$ the ground state energy.

\subsection*{Berry, 't Hooft, Polyakov and Us}

We now compute the Berry phase for this quantum mechanical system. The system is prepared in one of the ground states before the magnetic field $\vec{B}$  is adiabatically varied, traversing a closed loop in parameter space. At the end of this tour,  the ground state has undergone a $U(2)$ rotation, defined, as in \eqn{nonabb},  by the path ordered exponential of the Berry connection,
\begin{equation}
\vec{\cal A}_{ab} = i\langle b | \frac{\partial}{\partial\vec{B}}|\, a\rangle,\qquad a,b=1,2\ .
\end{equation}
To build some intuition, let's start with the diagonal components of the connection. Consider a large magnetic field $B\gg \hbar/m$, which localizes the spatial part of each wavefunction close to a pole, at $\theta=0$ or $\theta=\pi$. Here the ground state knows little about the rest of sphere and sees an  effective Hamiltonian of the form \eqn{ham}. This gives rise to a $U(1)$ Berry connection which is equal to that of a Dirac monopole, $\vec{\cal A}^{\ \rm Dirac}$. In fact, a simple computation reveals that the diagonal components are independent of the spatial wavefunctions for all values of $B$, and are given by
\be
\vec{\cal A}_{11}=\langle 1 | \frac{\partial}{\partial\vec{B}} |\,1\rangle &=& \langle \uparrow |  \frac{\partial}{\partial\vec{B}} \spup = \vec{A}{}^{\ \rm Dirac}\nonumber\ ,\\
\vec{\cal A}_{22}=\langle 2 | \frac{\partial}{\partial\vec{B}} | 2 \rangle &=& \langle \downarrow| \frac{\partial}{\partial\vec{B}} \spdn = -\vec{A}{}^{\ \rm Dirac} \ ,\label{diag}
\ee
In contrast, the off-diagonal terms describe the tunneling between the two different spin states, and depend on the spatial wavefunction of the particle $\psi$. They are,
\be\label{offdiag}
\vec{A}_{21}=\langle 2 | \frac{\partial}{\partial\vec{B}}|  1 \rangle &=& f(B)\  \langle \downarrow | \frac{\partial}{\partial\vec{B}} \spup \ ,\nonumber\\
\vec{A}_{12}=\langle 1 | \frac{\partial}{\partial\vec{B}} | 2 \rangle &=& f(B)\   \langle \uparrow | \frac{\partial}{\partial\vec{B}} \spdn\ ,
\ee
where the function $f(B)$ is the overlap,
\be f(B) = 2\pi\int_0^\pi \sin\theta\,d\theta\  \psi^\dagger(-\cos\theta)\,\psi(\cos\theta)\ . \label{f}
\ee
Without specific knowledge of the ground state wavefunction $\psi$, we are unable to compute explicitly the profile $f(B)$ of the non-Abelian Berry monopole. However, on general grounds, we know that $f(B)\rightarrow 0$ as $B\rightarrow \infty$ since the two spatial wavefunctions are localized at antipodal points on the sphere. In the opposite limit, $B=0$, the two spatial wavefunctions coincide and $f(0)=1$.

\para
The Dirac monopole connection $\vec{A}^{\ \rm Dirac}$ necessarily contains a singularity along a half-line,  known as the Dirac string. In the present context, this arises because it is not possible to globally define a basis of spin states $\spup$ and $\spdn$ for all values of $\vec{B}$. Therefore any explicit computation of the components of $\vec{\cal A}_{ab}$, using the basis shown in \eqn{diag} and \eqn{offdiag}, necessarily suffers from the Dirac string. However, there does exist a gauge in which the non-Abelian connection ${\cal A}$ is free from the Dirac string. To demonstrate this, one must first choose a $\vec{B}$ dependent basis for $\spup$ and $\spdn$, then rotate $\vec{\cal A}$ using a suitable singular gauge transformation. This was done, for example, in \cite{us22}. The result is the non-Abelian Berry connection which takes the rotationally covariant form,
\be
{\cal A}_\mu = \epsilon_{\mu\nu\rho} \frac{B_\nu \sigma^\rho}{2B^2}\left(1- f(B)  \right)\ .
\label{tpmon}\ee
This is the connection of a 't Hooft-Polyakov monopole. Note, firstly, that it is an $su(2)$ connection, rather than $u(2)$. Moreover, and most importantly, the asymptotic behaviour of $f(B)$ described above guarantees that, as $B\rightarrow \infty$, it reduces to the Dirac monopole for a $U(1) \subset SU(2)$. Yet the field strength is smooth at $B=0$.

\subsection{The BPS Monopole}

Any deformation of  the Hamiltonian \eqn{newham} that preserves the vacuum degeneracy will again lead to a 't Hooft-Polyakov monopole with a different profile function $f(B)$. For example, we may add a spin-blind potential to the Hamiltonian. Something special happens for the potential given by,
\be V(\theta) = \ft12  m B^2 \sin^2\theta \ .\label{v}
\ee
For this choice, the Schr\"odinger equation simplifies. The ground state energy is $E_0=0$, and it is a simple matter to find the exact wavefunctions. They are given by \eqn{gstate}, with
\be \psi(\cos\theta; B) = \left(\frac{Bm/\hbar}{2\pi \sinh(2Bm/\hbar)}\right)^{1/2}\,e^{(Bm/\hbar)\cos\theta}\ .
\label{exact}\ee
Equations \eqn{f} and \eqn{tpmon} then tell us the exact Berry connection for this system:
\begin{equation}\label{bps}
{\cal A}_\mu = \epsilon_{\mu\nu\rho} \frac{B_\nu \sigma^\rho}{2B^2}\left(1- \frac{2Bm/\hbar}{\sinh(2Bm/\hbar)}  \right)\ .
\end{equation}
Remarkably, this is exactly the profile function of the BPS monopole satisfying \eqn{bog}. We could ask whether the adjoint-valued Higgs field, $\phi$, also has a counterpart in our quantum mechanics. Indeed, it is given by the $su(2)$ valued expectation value,
\begin{equation}
\phi_{ab} =  \frac{2m}{\hbar}\,\langle b| \cos\theta | a\rangle\ . \label{higgs}
\end{equation}
Using the exact ground state \eqn{exact}, and after performing the gauge transformation to the rotationally covariant gauge described above, we find the scalar field profile,
\begin{equation}
\phi = \frac{B_i\sigma^i}{B^2}\left(\frac{2Bm}{\hbar}\coth\left(\frac{2Bm}{\hbar}\right)-1  \right)\ .
\end{equation}
which is precisely the form of the Higgs field for the BPS monopole solution of $SU(2)$ Yang-Mills Higgs theory \eqn{bog}. The magnetic field $\vec{B}$ plays the role of the spatial position, while the analog of the Higgs expectation value is $2m/\hbar$.

\para
The appearance of a Bogomolnyi equation, such as \eqn{bog}, usually hints at some underlying supersymmetry. Our model is no exception. Although the Hamiltonian with the potential \eqn{v} is not supersymmetric, it does turn out to be a consistent truncation from a supersymmetric theory. In the following section, we re-analyze the problem from this perspective.


\section{Supersymmetry and the Vacuum Bundle}\label{sec:susy}

Quantum mechanical sigma models with ${\cal N}=(2,2)$ supersymmetry have a target space ${\cal M}$ that admits a K\"ahler metric $g_{i\bar{j}}$. They arise from the dimensional reduction of ${\cal N}=1$ supersymmetric models in $d=3+1$ dimensions. The degrees of freedom consist of complex coordinates $z^i$ on ${\cal M}$, together with a pair of complex fermions $\psi_+^i$ and $\psi_-^i$ which are sections of the  tangent bundle of ${\cal M}$. The Hamiltonian is given by
\begin{equation}\label{eq:NLSM}
H = g^{i \bar j}\pi_i \bar\pi_{\bar j} + R_{i\bar j k \bar l}\,\psi^i_+\bar\psi^{\bar j}_+ \psi^k_- \bar\psi^{\bar l}_-\ ,
\end{equation}
where the momentum $\pi_{i} = g_{i\bar j}\dot{\bar{z}}^{\bar j}$ is defined in terms of the canonical momentum $p_{i}$ via
\be
\pi_{i} = p_{i} - ig_{k\bar j}\Gamma^{k}_{il}\left( \psi^l_+\bar\psi^{\bar j}_+ + \psi^l_-\bar\psi^{\bar j}_-\right)
\ee
The operators satisfy the (anti)-commutation relations,
\be\label{eq:relations}
\left[ \pi_i , z^j \right] = -i\delta_i^j \ \ \ ,\ \ \ \left[ \pi_i , \psi^j_\alpha \right] = i\Gamma^{j}_{\phantom{j}ik}\psi^k_\alpha\ \ \ , \ \ \
\left\{ \psi^i_\alpha , \bar\psi^{\bar j}_\beta \right\} = g^{i\bar j}\delta_{\alpha\beta}\ .
\ee
as well as the identity
\be\label{eq:riem}
\left[\pi_i,\bar\pi_{\bar j}  \right] = R_{i\bar j k\bar l}(\psi^k_+\bar\psi^{\bar l}_+ + \psi^k_-\bar\psi^{\bar l}_- )\ .
\ee
The only other non-zero (anti-)commutators are the conjugates of the above.The two complex supercharges of the theory are defined by $Q_\pm = \pi_i\psi_\pm^i$. These transform in the ${\bf 2}_{+1}$ representation of an $SU(2)_R\times U(1)_R$ R-symmetry.

\para
It was famously shown by Witten that the quantization of this model can be framed entirely in a geometric language \cite{witten}. We can view the fermions as creation and annihilation operators, and define a reference state $|\,\Omega\rangle$ such that $\psi_+^i|\,\Omega\rangle = \bar{\psi}_-^{\overline{i}}|\,\Omega\rangle = 0$. Acting with the creation operators $\bar{\psi}_+^{\bar{i}}$ and $\psi_-^i$ can then be thought of as wedging with forms,
\be
\bar \psi^{\bar j}_+ &\rightarrow\ d\bar z^{\bar j}\qquad , \qquad \psi^i_- \rightarrow\ dz^i\ .
\ee
In this way, the Hilbert space of states is identified with the space of square-integrable forms. The adjoint operators are
\be
\psi^i_+ \rightarrow\ g^{i\bar j}\imath_{\partial / \partial \bar z^{\bar j}} \qquad
, \qquad  \bar\psi_- \rightarrow\ g_{i\bar j}\imath_{\partial / \partial z^i}\ .
\ee
where $\imath_{v}$ denotes interior multiplication (contraction) with the vector $v$.
In this language, the supercharges are the Dolbeault operators on ${\cal M}$
\be
Q_-\rightarrow\ \partial \qquad, \qquad  Q_+  \rightarrow\ *\,\partial* \ .\ee
and the problem of determining the ground states of the theory, which satisfy $Q_\pm |a\rangle = \bar{Q}_\pm |a\rangle=0$, becomes the problem of determining the  Dolbeault cohomology of ${\cal M}$. The $SU(2)_R$ symmetry, under which the supercharges form a doublet, descends to the Lefschetz action on the cohomology \cite{C1,jose}.

\subsubsection*{The Mass Deformation}

There exists a massive deformation of the quantum mechanical sigma model that preserves all the supersymmetries \cite{agf}. One can view this in the following way: suppose that ${\cal M}$ has a holomorphic $U(1)$ isometry, with Killing vector $k^i$. We can consider weakly gauging this isometry. In quantum mechanics, the vector multiplet has three real scalars which form a triplet under the $SU(2)_R$ symmetry. Introducing background parameters for these scalars results in a potential over ${\cal M}$. The zeroes of this potential are the fixed points of $k$. These background parameters are usually denoted as a triplet of masses $\vec{m}$. However, to make contact with the results of Section 2, we will denote this triplet of vector multiplet parameter as $\vec{B}$. The mass-deformed Hamiltonian reads\footnote{In fact, this isn't the most general deformation. We may add a potential of this type associated to every mutually commuting holomorphic isometry of ${\cal M}$.}
\be
H = g^{i \bar j}\pi_i \bar\pi_{\bar j} +  g^{i\bar j}\,\bar k_i k_{\bar j}B^{2}+  i( \nabla_{i}k_{\bar j}) \bar \psi^{\bar j}\vec{B}\cdot  \vec{\sigma}\psi^i   + R_{i\bar j k \bar l}\,\psi^i_+\bar\psi^{\bar j}_+ \psi^k_- \bar\psi^{\bar l}_-\ .
\label{hmass}\ee
The supercharges are now given by
\be
Q_\alpha &= &\pi_i \psi^i_\alpha + \bar k_i ( \vec{B} \cdot \vec{\sigma})_{\alpha\beta} \psi^i_\beta\ , \qquad\alpha, \beta = \pm 1
\ee
Using the Killing equation for $k$, it can be shown that these obey the superalgebra
\be
\left\{ Q_\alpha,Q_\beta \right\}& =& \left\{ \overline{Q}_\alpha,\overline{Q}_\beta \right\}=0\ ,\nonumber\\
\left\{ Q_\alpha,\overline{Q}_\beta \right\} &= &\delta_{\alpha\beta}\,H + \vec{B} \cdot \vec{\sigma}_{\alpha\beta}\,{\cal Z} \ ,
\ee
with central charge ${\cal Z}= (k^{i} \pi_i + \bar k^{\bar j} \bar\pi_{\bar j}) +i(\nabla_{i}k_{\bar j})\bar\psi^{\bar j}_{\alpha}\psi^{i}_{\alpha}$.

\para
The Witten index ensures that, for compact target spaces, the number of vacuum states (counted with sign) remains unchanged under the deformation $\vec{B}$. However, the ground state wavefunctions do change. Once again translating to the geometric language, the supercharges become,
\be
\left( \begin{array}{c} Q_+ \\ Q_- \end{array} \right) \rightarrow  \left( \begin{array}{c} *\, \partial * \\ \partial \end{array} \right) + (\vec{B}\cdot\vec{\sigma})\,
\left(\begin{array}{c}  \imath_{\bar k} \\ \, *\, \imath_{\bar k}* \end{array} \right)\ .
\ee
For the specific choice of $B_3=0$ (so that $\vec{B}\cdot\vec{\sigma}$ is off-diagonal), the supercharges $Q_\pm$ give rise to the equivariant cohomology \cite{lag} of ${\cal M}$ with respect to the $U(1)$ action  generated by $k$. The operators above provide an $SU(2)_R$ covariant version of this. The ground states of the quantum mechanics define a {\it vacuum bundle} over ${\bf R}^3$,  the space of parameters $\vec{B}$. Our interest is in the way this vacuum bundle is fibered over ${\bf R}^3$, and the associated Berry connection.

\subsection{The ${\bf CP}^1$ Sigma-Model}

We first look at the ${\bf CP}^1$ sigma-model. We will find that the ground states are those of Section 2, resulting in the BPS monopole as the Berry phase. We previously studied this system in \cite{us22} from the perspective of the gauged linear sigma-model. There, we showed that the Berry phase was an $SU(2)$ 't Hooft-Polyakov monopole and computed the first leading order instanton contribution to the monopole profile. Below we confirm that the exact results of this paper agree with our earlier analysis.

\para
The metric on ${\bf CP}^1$ is given by,
\be ds^2 = \frac{m}{2}\frac{dzd\bar{z}}{(1+|z|^2)^2}\ , \ee
where $m$ is the K\"ahler class of the manifold. The kinetic terms for the sigma model coincide with those of \eqn{newham} under the identification of K\"ahler class with particle mass, together with $\hbar=1$ and $z=\tan(\theta/2)e^{i\phi}$. Furthermore, the potential term in \eqn{hmass} coincides with \eqn{v} for the choice of Killing vector $k=iz(\partial/\partial z) -i\bar{z}(\partial/\partial \bar{z}) = \partial/\partial \phi$.

\para
However, the supersymmetric quantum mechanics is not identical to the system discussed in Section 2. Quantizing the fermions gives rise to Hilbert space of dimension 4, corresponding to the $(p,q)$-forms on ${\bf CP}^1$. It is simple to check that, for all values of the parameters $\vec{B}$, the ground states live in the even cohomology, i.e. the zero-form and the top-form. (Or, in the language of fermion creation operators, they only involve the states $|\,\Omega\rangle$ and $\psi_-\bar{\psi}_+|\,\Omega\rangle$). If we are interested only in the properties of these ground states (and we are!) then we may restrict attention to these two sectors. The differential equations governing the Berry phase may then be viewed as arising from an effective two-state system moving on the sphere. The resulting Hamiltonian is precisely that given in \eqn{newham}, together with the specific potential \eqn{v}. Therefore, the Berry connection of the ${\bf CP}^1$ sigma-model is given by the BPS monopole \eqn{bps}.

\para
It is illustrative to expand the profile function $f(Bm)= 2Bm/\sinh(2Bm)$,
\begin{equation}
f(B) = 4Bm e^{-2Bm}\left(1+ {\cal O}(e^{-4Bm})\right)\ .
\end{equation}
This leading order contribution to the monopole profile was shown in \cite{us22} to arise from a BPS instanton in the quantum mechanics. The exact result above agrees with the explicit instanton computation in \cite{us22}\footnote{To compare with the conventions of \cite{us22}, we need the dictionary $m\rightarrow r$ and $\vec{B}\rightarrow \vec{m}$. However,~importantly,~$4\rightarrow 4$.}.
The higher order contributions arise from instanton-anti-instanton pairs, bouncing back and forth between the two vacua. It is amusing that the long-known form of the profile function of the BPS monopole can be interpreted an instanton expansion in supersymmetric quantum mechanics.

\para
Finally, we mention that the construction of the Higgs field \eqn{higgs} has a rather natural mathematical meaning. As explained at the beginning of this section, the potential over the target space manifold ${\cal M}$ is related to a holomorphic Killing vector $k$. For K\"ahler manifolds, this Killing vector arises from a {\it moment map} $\mu$. (In the physics literature these maps are also known as Killing potentials). This is a function over ${\cal M}$, defined such that $d\mu = \imath_{k}\omega$, where $\omega$ is the K\"ahler form. For the case of ${\CP}^1$ and the vector field $k=\partial/\partial\phi$, the moment map is given by  $\mu = \cos\theta$. The Higgs field $\phi$ is simply the expectation value of this moment map.
\be \phi_{ab} = \frac{m}{2}\langle b\, | \mu |\, a\rangle \ .\ee
The way in which the monopole and Higgs field arise in this context is reminiscent of  Nahm's construction \cite{Nahm:1979yw}, with the coordinate $\theta \in [0,\pi)$ playing the role of the interval in Nahm's story.

\subsection{A Conjecture for $\mathbf{CP}^N$}

Nahm's construction also reproduces multi-monopole solutions and monopoles in higher rank gauge groups. It is interesting to ask whether the Berry phases associated to sigma-models with other target spaces are also given by BPS monopoles. We conjecture that this is indeed the case, at least for the ${\bf CP}^{N}$ target spaces.

\para
The ${\bf CP}^N$ supersymmetric sigma model has $N+1$ ground states, and these are protected by the Witten index as we turn on a potential. In fact, there are $N$ such potentials that we could turn on, corresponding to $N$ orthogonal, holomorphic, $U(1)$  Killing vectors on ${\bf CP}^N$. Denote the moment maps as $\mu_m$, $m=1,\ldots,N$. We turn on a potential associated to the linear combination  $\zeta=t^m\mu_m$ for fixed $t^m$. As before, the Hamiltonian depends on three parameters $\vec{B}$, which govern the overall scale and $SU(2)_R$ orientation of the potential. The expectation value of the moment map defines an $SU(N+1)$ adjoint-valued  Higgs field, with asymptotic behaviour
\be \phi_{ab}=\frac{m}{2}\langle b|\zeta|\,a\rangle\ \longrightarrow \ \frac{m}{2}t^mH^m_{ab} \qquad  {\rm as}\ B\rightarrow \infty\ .\label{assy}\ee
%
%
where $H^m$ are the Cartan generators of $SU(N+1)$. For generic choices of $t^m$, this Higgs expectation value breaks the gauge group to the Cartan subalgebra: $SU(N+1)\rightarrow U(1)^N$.

\para
As we vary $\vec{B}$, the mixing of the ground states is determined by the $SU(N+1)$ Berry connection over ${\bf R}^3$. It is not hard to see that, asymptotically, the Berry connection looks like a single Dirac monopole sitting in each element of the Cartan subalgebra $U(1)^N\subset SU(N+1)$; that is, a monopole with magnetic charge $(1,1,\ldots,1)$. It is natural to conjecture that the full Berry phase in this case is once again given by the BPS monopole, satisfying \eqn{bog}.

\para
The $SU(N)$ Berry connection described above has $SU(2)_R$ rotational invariance. However, the $(1,1,\ldots,1)$ monopole in $SU(N)$ has a $4(N-1)$ dimensional moduli space (ignoring the translational degrees of freedom) \cite{lwy}. There is a unique point on this moduli space corresponding to an $SU(2)_R$ rotationally invariant monopole. At other points on the moduli space, the constituent monopoles have separated and the configuration is no longer rotationally invariant. In fact, there is also a natural guess for how these configurations arise as a Berry connection. We may alter the ${\bf CP}^N$ model, preserving supersymmetry, by turning on a second potential with moment map orthogonal to $\zeta$. (Meaning that the associated Killing vectors are orthogonal). We leave this potential fixed, while varying the coefficients $\vec{B}$ that govern the potential $\zeta$. In this way, we can generate a family of $SU(N)$ Berry connections over ${\bf R}^3$. The dimension of the moduli space of Berry connections is equal to  the dimension of the monopole moduli space. (Strictly speaking, we generate a $3(N-1)$-dimensional space in this manner. However, turning on chemical potentials, associated to the $A_0$ component of a background vector multiplet, generates the remaining moduli).

\para
It would be interesting to prove the speculations in this section, and to understand if ${\cal N}=(2,2)$  quantum mechanical sigma-models with other target spaces also have BPS monopoles as their Berry connections\footnote{{\bf Note added:} They do! The Berry phase for any quantum mechanical 
sigma-model, arising as $\vec{m}$ is varied, obeys the Bogomolnyi equation \eqn{bog}. A proof of this was given in \cite{holonomy}.}. There is also an interesting open question regarding the connection to the tt* equations \cite{C1,C2,tt*}. The tt* equations apply to the variation of parameters that live in background chiral multiplets. Moveover, they hold both in quantum mechanics and in $d=1+1$ dimensional theories. In contrast, the discussion in this paper holds only in quantum mechanics, since only there does the vector multiplet contain three real scalars. However, mirror symmetry should provide a connection. In two dimensions, vector multiplet scalars can be packaged into twisted chiral multiplets which, in turn, are related to chiral multiplets through mirror symmetry. This hints at a deeper relationship between the tt* equations and the Bogomolnyi equations.

\newpage

\acknowledgments{We would like to thank Philip Argyres, Jan de Boer, Nick Dorey,  Koji Hashimoto, Kentaro Hori, 
Kyriakos Papadodimas,  Chris Pedder, Joseph Samuel, Alfred Shapere and Cumrun Vafa for useful conversations. This work is supported in part by funds provided by the U.S. Department of Energy (D.O.E.) under cooperative research agreement DEFG02-05ER41360. J.S. is supported by NSF grant PHY-0600465 and thanks the members of CTP at MIT, and especially Dan Freedman, for hospitality. D.T. is supported by the Royal Society.}

\end{document}